\documentclass[%
reprint,
superscriptaddress,
showpacs,preprintnumbers,
prl,
]{revtex4-1}

\usepackage{graphicx}
\usepackage{dcolumn}
\usepackage{bm}

\usepackage{color} 

\begin{document}

\preprint{APS/123-QED}

\title{Irregular Oscillatory-Patterns in the Early-Time Region of Coherent Phonon Generation in Silicon }

\author{Yohei Watanabe}
\affiliation{Doctoral Program in Materials Science, Graduate School of Pure and Applied Sciences, University of Tsukuba, Tsukuba, Ibaraki 305-8573, Japan}
\author{Ken-ichi Hino}
\email{hino@ims.tsukuba.ac.jp}
\affiliation{Division of Materials Science, Faculty of Pure and Applied Sciences, University of Tsukuba, Tsukuba 305-8573, Japan}
\affiliation{Center for Computational Sciences, University of Tsukuba, Tsukuba 305-8577, Japan}
\author{Muneaki Hase}
\affiliation{Division of Applied Physics, Faculty of Pure and Applied Sciences, University of Tsukuba, Tsukuba 305-8573, Japan}
\author{Nobuya Maeshima}
\affiliation{Center for Computational Sciences, University of Tsukuba, Tsukuba 305-8577, Japan}
\affiliation{Division of Materials Science, Faculty of Pure and Applied Sciences, University of Tsukuba, Tsukuba 305-8573, Japan}

\date{\today}

\begin{abstract}
Coherent phonon (CP) generation in an undoped Si crystal is theoretically investigated to shed light on unexplored quantum-mechanical effects in the early-time region immediately after the irradiation of ultrashort laser pulse.
One examines time signals attributed to an induced charge density of an ionic core, placing the focus on the effects of the Rabi frequency $\Omega_{0cv}$ on the signals; this frequency corresponds to the peak electric-field of the pulse.
It is found that at specific $\Omega_{0cv}$'s where the energy of plasmon caused by photoexcited carriers coincides with the longitudinal-optical phonon energy,
the energetically {\it resonant } interaction between these two modes leads to striking anticrossings, revealing irregular oscillations with anomalously enhanced amplitudes in the observed time signals. 
Also, the oscillatory pattern is subject to the Rabi flopping of the excited carrier density that is controlled by $\Omega_{0cv}$.
These findings show that the early-time region is enriched with quantum-mechanical effects inherent in the CP generation, though experimental signals are more or less masked by the so-called coherent artifact due to nonlinear optical effects.

\end{abstract}

\maketitle



Coherent phonon (CP) generation is one of the representative ultrafast phenomena induced by an ultrashort pulse laser, and a great number of studies on this have been reported in diverse fields of physics and chemistry \cite{silvestri}.
In condensed-matter systems, CPs are observed in various bulk materials and heterostructures
such as semiconductors, semimetals, superconductors, and so on~\cite{kuznetsov1,cp,hase1,hase2}.
The exploration of the CP generation mechanism has been targeted for these experimental studies, in which observed signals are examined based on the classical model following a damped forced harmonic oscillation.
In particular, an initial phase built in the oscillator
is considered as a key quantity for the generation mechanism.
Most of theoretical studies are devoted to analyzing this phase in either phenomenological or semi-classical manner~\cite{initphase}.
However, such an approach falls into difficulty of revealing quantum-mechanical effects inherent in the CP generation.
These effects are dominant in the initial stage of CP dynamics in which a great number of photoexcited carriers still stay in nonequilibrium states before being relaxed; hereafter, this stage is termed as the early-time region (ETR).
In fact, intricate signals observed in ETR have been interpreted as coherent artifact (CA) due to nonlinear optical interference, making it difficult to extract information on the intrinsic dynamics of investigated materials~\cite{ca}.
It is just the transient Fano resonance that has been studied as a quantum-mechanical effect inherent in the CP generation; the detail of it has been brought to light recently~\cite{hase1,watanabe,riffe2}.

The aim of this Letter is to explore quantum-mechanical effects still hidden in ETR of the CP generation in undoped Si based on the polaronic-quasiparticle (PQ) model developed by the authors \cite{watanabe}.
Here, the high-density carriers generated by intense ultrashort-pulse lasers 
fulfill a vital role, leading to a strong interaction with a longitudinal optical (LO) phonon.
In particular, a collective excitation mode (plasmon) resulting from the carriers is focused, and its effects on the amplitude and initial phase of CP signal are evaluated.
Atomic units are used throughout unless otherwise stated.

The total Hamiltonian of this system, represented by $\hat{H}(t)$, consists of an electron Hamiltonian $\hat{H}_e$ including a Coulomb interaction between electrons, an electron-laser interaction $\hat{H}^\prime(t)$ at time $t$, an LO-phonon Hamiltonian with its energy dispersion $\omega_{\boldsymbol{q}}$ at Bloch momentum $\boldsymbol{q}$, and a deformation-potential interaction between electron and LO-phonon $\hat{H}_{e-p}$.
The external-laser electric-field is given by
\(
F(t)=f(t)\cos\omega_0 t,
\)
where $\omega_0$ represents the center frequency, and $f(t)$ is a Gaussian-shaped pulse-envelope function 
with a peak amplitude $F_0$ and a temporal width $\tau_L$ --- the full width at half maximum --- satisfying $\tau_L \ll T_{\boldsymbol{q}} \equiv 2\pi /\omega_{\boldsymbol{q}}$.
The magnitude of $\hat{H}^\prime(t)$ is determined by the Rabi frequency $\Omega_{0cv}$ given by the product of $F_0$ and the electronic dipole moment between $\Gamma$-points of conduction $(c)$ and valence $(v)$ bands.
The magnitude of $\hat{H}_{e-p}$ is determined by a coupling constant $g_{b\boldsymbol{q}}$ of $b$-band electron $(b=c,v)$ with the LO phonon.
Hereafter, creation and annihilation operators of electron in $b$-band with Bloch momentum $\boldsymbol{k}$ are represented by $a^\dagger_{b\boldsymbol{k}}$ and $a_{b\boldsymbol{k}}$, respectively, while those of LO-phonon with momentum $\boldsymbol{q}$ are represented by $c_{\boldsymbol{q}}^\dagger$ and $c_{\boldsymbol{q}}$.

The nonequilibrium carrier dynamics driven by $F(t)$ is governed by the time-evolution of a composite operator, $A_{\boldsymbol{q}}^\dagger(\boldsymbol{k}bb') = a^\dagger_{b\boldsymbol{k+q}} a_{b^\prime\boldsymbol{k}}$,
representing a carrier density matrix for the transition from $b'$-band to $b$-band with an anisotropic momentum distribution.
The transferred momentum $\boldsymbol{q}$ is quite small, but finite in the CP generation, namely, $|\boldsymbol{q}| \neq 0$, and the $t$-dependence of this operator is omitted for the sake of simplicity.
The problem of concern is made tractable by applying the rotating wave approximation for the equation of motion of $A_{\boldsymbol{q}}^\dagger(\boldsymbol{k}bb')$, leading to the equation of motion of a transformed operator $\bar{A}_{\boldsymbol{q}}^\dagger(\boldsymbol{k}bb')=
A_{\boldsymbol{q}}^\dagger(\boldsymbol{k}bb')e^{-i \bar{\omega}_{bb'}t}$,
where $\bar{\omega}_{cv}=\omega_0$, $\bar{\omega}_{vc}=-\omega_0$, and $\bar{\omega}_{bb}=0$.
In fact, it is illustrated that when $t > \tau_L/2$, this equation
is dealt with based on the adiabatic approximation with respect to $t$~\cite{watanabe,SM,comm1}.
Thus, 
the carriers even in nonequilibrium can be classified into two modes of collective excitation and individual excitation in an adiabatic sense.
The former mode corresponds to a plasmon, and this is obtained by applying the bosonization procedure for an intraband contribution of $\hat{H}_e$; an exciton mode is neglected because of little contributions here.
The resulting creation operator of plasmon, represented as $B^\dagger_{\boldsymbol{q}}$, is described by a linear combination of the intraband density matrices
$A_{\boldsymbol{q}}^\dagger(\boldsymbol{k}bb)$'s \cite{haug}.
The creation operator of the individual excitation mode is represented by $A_{\boldsymbol{q}}^\dagger(\boldsymbol{k}b\bar{b})$ with $\bar{b} \not = b$, where
the intraband contribution is neglected because 
$|\boldsymbol{q}| \approx 0$.

Here, for the sake of convenience, the following notations are introduced: $\hat{O}_{\boldsymbol{q}ph} = c_{\boldsymbol{q}}^\dagger$, $\hat{O}_{\boldsymbol{q}(\boldsymbol{k}b\bar{b})} = \bar{A}_{\boldsymbol{q}}^\dagger (\boldsymbol{k}b\bar{b})$, and $\hat{O}_{\boldsymbol{q}pl} = B_{\boldsymbol{q}}^\dagger$.
Thus, the equation of motion of $\hat{O}_{\boldsymbol{q}j}$ is given by
\( -i(d/dt + \gamma_{\boldsymbol{q}j} )\:\hat{O}_{\boldsymbol{q}j}=[\hat{H}(t), \hat{O}_{\boldsymbol{q}j}]=\sum_{j'}\hat{O}_{\boldsymbol{q}j'}\bar{Z}_{\boldsymbol{q}j'j}
\)
with $\bar{Z}_{\boldsymbol{q}}$ as a non-Hermitian matrix and $\gamma_{\boldsymbol{q}j}$ as a damping constant of the $j$th mode; $j, j'=\{ph, (\boldsymbol{k}b\bar{b})\cdots, pl \}$.
Now, the PQ operator is introduced by
\(
P_{\boldsymbol{q}j}^\dagger
=\sum_{j'} \hat{O}_{\boldsymbol{q}j'} V_{\boldsymbol{q}j'j}^{R}
\).
Here, the left and right eigenvalue problems of $\bar{Z}_{\boldsymbol{q}}$ given by $V_{\boldsymbol{q}j}^{L\dagger} \bar{Z}_{\boldsymbol{q}}= {E}_{\boldsymbol{q}j}V_{\boldsymbol{q}j}^{L\dagger}$ and $\bar{Z}_{\boldsymbol{q}}V_{\boldsymbol{q}j}^{R}=V_{\boldsymbol{q}j}^{R} {E}_{\boldsymbol{q}j}$,
respectively, are solved to obtain the $j$th adiabatic eigenvalue ${E}_{\boldsymbol{q}j}$ and the corresponding biorthogonal set of eigenvectors $\{ V_{\boldsymbol{q}j}^{L}, V_{\boldsymbol{q}j}^{R} \}$.
The time-evolution of $P_{\boldsymbol{q}j}^\dagger$ is obtained by solving the associated Heisenberg equation.
Thus, the retarded phonon Green function 
\(
D_{\boldsymbol{q}}^R(t,t')
=
\bar{D}_{\boldsymbol{q}}^R(t,t') + \left[ \bar{D}_{\boldsymbol{-q}}^R(t,t') \right]^*
\), 
where
$\bar{D}_{\boldsymbol{q}}^R(t,t')=-i \langle [c_{\boldsymbol{q}}(t), c_{\boldsymbol{q}}^\dagger (t')] \rangle
\theta (t-t')$, is readily expressed
in terms of the PQ operators, since $c_{\boldsymbol{q}}^\dagger = \sum_{j} P_{\boldsymbol{q}j}^\dagger V_{\boldsymbol{q}j,ph}^{L\dagger}$.
For more detail, consult Ref.~\cite{SM}.

Following the linear response theory, $D_{\boldsymbol{q}}^R(t,t')$ shows an induced charge density of ionic cores probed at time $t'$ by a weak test potential with a delta-function form $\delta(t')$.
The induced charge density 
due to CP generation is given by
$Q_{\boldsymbol{q}}(\tau)\equiv D_{\boldsymbol{q}}^R(\tau+t',t')-D_{\boldsymbol{q}}^{R(0)}(\tau+t',t')$ aside from an unimportant proportional constant with $\tau=t-t' \ge 0$. Here, the free phonon Green function $D_{\boldsymbol{q}}^{R(0)}(\tau+t',t')$ without the pump field 
is subtracted because this contributes to the incoherent phonon generation.
Thus, $Q_{\boldsymbol{q}}(\tau)$ shows an oscillatory pattern of the CP precisely after the probe time $t'$.
Hereafter, the time of $t'=0$ is exclusively concerned.
To this end, $Q_{\boldsymbol{q}}(\tau)$ is rewritten as
\(
Q_{\boldsymbol{q}}(\tau) = A_{\boldsymbol{q}}(\tau) \cos \left[\omega_{\boldsymbol{q}}\tau+\Theta_{\boldsymbol{q}}(\tau) \right]
\), 
where $\Theta_{\boldsymbol{q}}(\tau) $ and $A_{\boldsymbol{q}}(\tau)$ represent a renormalized phase modulus $\pi$ and a transitory amplitude at $\tau$, respectively.
In the large-$\tau$ limit, $Q_{\boldsymbol{q}}(\tau)$ becomes a damped harmonics with $A_{\boldsymbol{q}}(\tau) \rightarrow A^0_{\boldsymbol{q}} \:e^{-\gamma_{\boldsymbol{q}ph}\tau}$ and $\Theta_{\boldsymbol{q}}(\tau) \rightarrow \theta_{\boldsymbol{q}}$, where the asymptotic amplitude $A^0_{\boldsymbol{q}}$ and the initial phase $\theta_{\boldsymbol{q}}$ are constants, and $\gamma_{\boldsymbol{q}ph}$ arises from phonon anharmonicity.

\begin{figure}[tb]
\begin{center}
\includegraphics[width=7.0cm,clip]{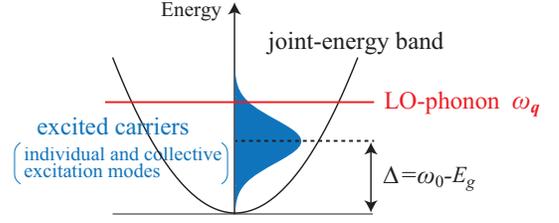}
\caption{The scheme of the CP generation dynamics. $\Delta = \omega_0 - E_g$ represents the detuning with $E_g$ as the direct band gap of Si at $\Gamma$ point. For more detail, consult the text.}
\label{fig1}
\end{center}
\end{figure}

\begin{figure}[tb]
\begin{center}
\includegraphics[width=7.0cm,clip]{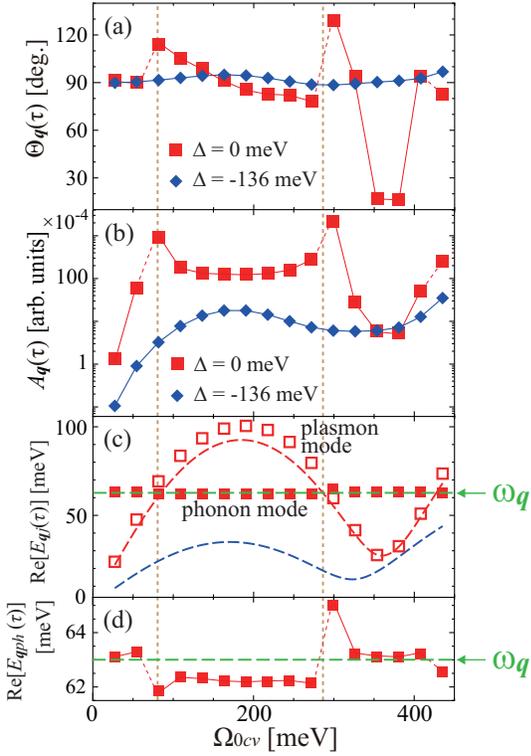}
\caption{(a) $\Theta_{\boldsymbol{q}}(\tau)$ and (b) $A_{\boldsymbol{q}}(\tau)$ as a function of $\Omega_{0cv}$ for $\Delta=\Delta_0$ (red square) and $\Delta_-$ (blue diamond).
(c) The real parts of the adiabatic energy $E_{\boldsymbol{q}j}(\tau)$ as a function of $\Omega_{0cv}$ for $j=ph, pl$; the eigenvalues mostly dominated by the phonon and plasmon modes are represented by filled and open red squares, respectively.
$\omega_{\boldsymbol{q}pl}(\tau)$ for $\Delta=\Delta_0$ (red dash line) and $\Delta_-$ (blue dash line) are also included with $\omega_{\boldsymbol{q}}=63$ meV (green dash line).
(d) The enlarged view of panel (c) around ${\rm Re}\:[E_{\boldsymbol{q}ph}(\tau)]=\omega_{\boldsymbol{q}}$.
In all panels, $\tau=20\:{\rm fs}$, and the positions of $\Omega_{0cv}^{(C1)}$ and $\Omega_{0cv}^{(C2)}$ are denoted by vertical brown dotted lines.
The calculated values in panels (a), (b), and (d) are connected by solid
and dashed lines in order to aid the presentation.
}
\label{fig2}
\end{center}
\end{figure}

Figure~\ref{fig1} diagrams the scheme of the CP generation dynamics showing that carriers are excited by the pump pulse with detuning $\Delta$ to form the energy distribution in the joint-band energy dispersion and the energy of LO-phonon $\omega_{\boldsymbol{q}}$ is partially overlapped with it; $\Delta \equiv \omega_0 - E_g$ with $E_g$ as the direct band gap at $\Gamma$ point.
Figures~\ref{fig2}(a) and \ref{fig2}(b) show the calculated results of $\Theta_{\boldsymbol{q}}(\tau)$ and $A_{\boldsymbol{q}}(\tau)$ at $\tau=20$ fs in ETR as a function of $\Omega_{0cv}$, respectively. 
Here, material parameters given in Ref.~\cite{watanabe} are employed, and $\tau_L=10$ fs that corresponds to spectral width of the laser of 370 meV.
It is seen that both $\Theta_{\boldsymbol{q}}(\tau)$ and $A_{\boldsymbol{q}}(\tau)$ for $\Delta = \Delta_0 \equiv 0$ meV change in an irregular manner with cusp structures at $\Omega_{0cv}= \Omega^{(C1)}_{0cv}\equiv 82$ meV and $\Omega^{(C2)}_{0cv}\equiv 286$ meV, and the envelopes of both functions change steeply around $\Omega_{0cv}=350$ meV. 
This contrasts with the behavior of $\Theta_{\boldsymbol{q}}(\tau)$ and $A_{\boldsymbol{q}}(\tau)$ for $\Delta = \Delta_-\equiv -136$ meV, showing even more moderate alteration over $\Omega_{0cv}$.

To deepen the understanding of this result, the real parts of adiabatic energy
$E_{\boldsymbol{q}j}(\tau)$ as a function of $\Omega_{0cv}$ are evaluated, as shown in Fig.~\ref{fig2}(c), where filled and open red squares represent the eigenvalues mostly dominated by phonon $(j=ph)$ and plasmon $(j=pl)$, respectively.
The plasma frequency $\omega_{\boldsymbol{q}pl}(\tau)$ in proportion to the square root of the whole excited carrier density is also included for $\Delta = \Delta_0$ and $\Delta_-$ in addition to $\omega_{\boldsymbol{q}}$.
It is evident that $\omega_{\boldsymbol{q}pl}(\tau)$ for $\Delta = \Delta_0$ coincides with $\omega_{\boldsymbol{q}}$ at $\Omega_{0cv}=\Omega^{(C1)}_{0cv}$ and $\Omega^{(C2)}_{0cv}$, leading to anticrossings between ${\rm Re}\:E_{\boldsymbol{q}ph}(\tau)$ and ${\rm Re}\:E_{\boldsymbol{q}pl}(\tau)$.
The detail of this phenomenon is shown in the enlarged figure of Fig.~\ref{fig2}(d).
The difference of ${\rm Re}\:E_{\boldsymbol{q}ph}(\tau)$ from $\omega_{\boldsymbol{q}}$ represents the self-energy resulting mostly from the interaction of the phonon with the plasmon; the contribution of the individual excitation mode 
is found negligibly small.
The sharp change of the self-energy for $\Delta = \Delta_0$ seen at $\Omega_{0cv}=\Omega^{(C1)}_{0cv}$ and $\Omega^{(C2)}_{0cv}$ is in harmony with the manifestation of anomalous cusp structures stated above.
Thus, it is concluded that the irregularity revealed here in both $\Theta_{\boldsymbol{q}}(\tau)$ and $A_{\boldsymbol{q}}(\tau)$ is unequivocally due to the anticrossings caused by the energetically {\it resonant} interaction of the phonon with the plasmon induced by laser pulse irradiation. 
According to Fig.~\ref{fig2}(d), the plasmon-phonon interaction remains effective in the range of $\Omega_{0cv}$ of $[\Omega^{(C1)}_{0cv}, \Omega^{(C2)}_{0cv}]$.
On the other hand, there is not such irregularity for $\Delta = \Delta_-$ because of 
$\omega_{\boldsymbol{q}} >\omega_{\boldsymbol{q}pl}(\tau)$ within the concerned range of $\Omega_{0cv}$.

Figure~\ref{fig2}(c) also shows that $\omega_{\boldsymbol{q}pl}(\tau)$ oscillates with an approximate period of 350 meV.
This is due to the interband Rabi flopping of the excited carriers that ends at $t \approx \tau_L/2$, since the rough estimate of $2\pi$-pulse is $\Omega_{0cv}=\Omega^{(2\pi)}_{0cv}\equiv 388$ meV besides the Coulomb correction; that of $\pi$-pulse is $\Omega^{(\pi)}_{0cv}\equiv \Omega^{(2\pi)}_{0cv}/2$.
Thus, the obvious changes of $\Theta_{\boldsymbol{q}}(\tau)$ and $A_{\boldsymbol{q}}(\tau)$ around $\Omega^{(2\pi)}_{0cv}$ for $\Delta = \Delta_0$ arise from the Rabi oscillation [see Figs.~\ref{fig2}(a) and \ref{fig2}(b)].

\begin{figure}[tb]
\begin{center}
\includegraphics[width=7.0cm,clip]{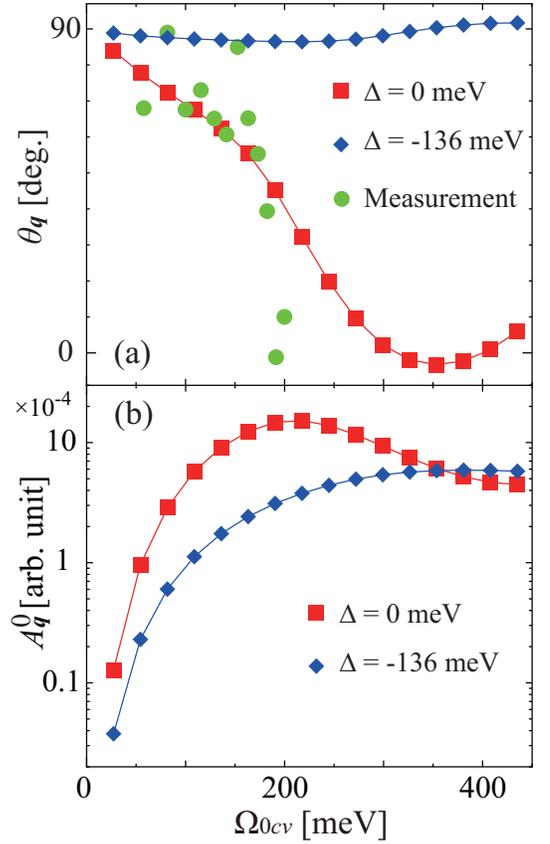}
\caption{(a) $\theta_{\boldsymbol{q}}$ and (b) $A^0_{\boldsymbol{q}}$ as a function of $\Omega_{0cv}$ for $\Delta=\Delta_0$ (red square) and $\Delta_-$ (blue diamond) along with the experimental data of $\theta_{\boldsymbol{q}}$ (green circle) \cite{hase2}.
}
\label{fig3}
\end{center}
\end{figure}

Figures~\ref{fig3}(a) and \ref{fig3}(b) show the calculated results of $\theta_{\boldsymbol{q}}$ and $A^0_{\boldsymbol{q}}$ as a function of $\Omega_{0cv}$, respectively.
The Rabi-oscillatory patterns 
still remain in both of $\theta_{\boldsymbol{q}}$ and $A^0_{\boldsymbol{q}}$ for $\Delta = \Delta_0$ around $\Omega_{0cv}=\Omega^{(2\pi)}_{0cv}$, while the cusp structure disappears since 
the plasmon-phonon interaction vanishes out of ETR.
The experimental data are also included in Fig.~\ref{fig3}(a), showing the dependence of $\theta_{\boldsymbol{q}}$ on the pump fluence~\cite{hase2}. 
With the increase of the fluence, $\theta_{\boldsymbol{q}}$ changes from $90^\circ$ to the vicinity of $0^\circ$, which is in accordance with the calculated results for $\Delta = \Delta_0$.

\begin{figure}[tb]
\begin{center}
\includegraphics[width=7.0cm,clip]{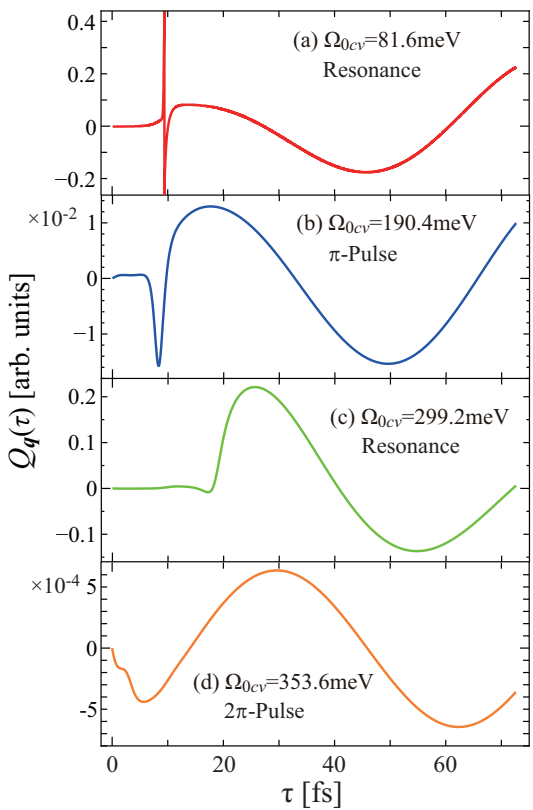}
\caption{$Q_{\boldsymbol{q}}(\tau)$ as a function of $\tau$ in ETR at the specific $\Omega_{0cv}$'s given in panels (a)-(d) with $\Delta=\Delta_0$.
}
\label{fig4}
\end{center}
\end{figure}

Figures~\ref{fig4}(a)-\ref{fig4}(d) show the alteration of $Q_{\boldsymbol{q}}(\tau)$ as a function of $\tau$ in ETR for $\Delta=\Delta_0$ at the four values of $\Omega_{0cv}$, namely, 
$\Omega^{(C1)\prime}_{0cv}\equiv 81.6$ meV, $\Omega^{(\pi)\prime}_{0cv}\equiv 190.4$ meV, $\Omega^{(C2)\prime}_{0cv}\equiv 299.2$ meV, and $\Omega^{(2\pi)\prime}_{0cv}\equiv 353.6$ meV, in the vicinity of
$\Omega^{(C1)}_{0cv}$, $\Omega^{(\pi)}_{0cv}$, $\Omega^{(C2)}_{0cv}$, and $\Omega^{(2\pi)}_{0cv}$, respectively \cite{comm1}.
Here, the number of excited carriers is mostly maximized and minimized at $\Omega^{(\pi)\prime}_{0cv}$ and $\Omega^{(2\pi)\prime}_{0cv}$, respectively.
At $\Omega_{0cv} = \Omega^{(C1)\prime}_{0cv}$ and $\Omega^{(C2)\prime}_{0cv}$, it is seen that
$Q_{\boldsymbol{q}}(\tau)$'s show marked irregularity due to the resonant interaction of the plasmon with the phonon from a simple harmonics with a period $T_{\boldsymbol{q}}=66$ fs.
In particular, it is noted that the transitory amplitudes $A_{\boldsymbol{q}}(\tau)$ at both $\Omega_{0cv}$'s are about ten times greater than that at $\Omega_{0cv}=\Omega^{(\pi)\prime}_{0cv}$,
while in contrast, the asymptotic amplitudes $A^0_{\boldsymbol{q}}$ at the former $\Omega_{0cv}$'s are several times less than that at the latter one; see Fig.~\ref{fig3}(b).
Further, the renormalized phase $\Theta_{\boldsymbol{q}}(\tau)$ show an anomalous oscillatory pattern, especially, at $\Omega_{0cv}=\Omega^{(C1)\prime}_{0cv}$, where this phase changes rapidly over $2\pi$ around $\tau=10$ fs due presumably to the manifestation of strong anticrossing. 
Also, it is seen that $Q_{\boldsymbol{q}}(\tau)$ is modulated by the maximized carrier inversion at $\Omega_{0cv}=\Omega^{(\pi)\prime}_{0cv}$.
Unlike it, at $\Omega_{0cv}=\Omega^{(2\pi)\prime}_{0cv}$, both $\Theta_{\boldsymbol{q}}(\tau)$ and $A_{\boldsymbol{q}}(\tau)$ remain almost unaltered, and are gradually close to the respective asymptotes; the resulting $Q_{\boldsymbol{q}}(\tau)$ just shows a damped harmonic oscillation almost in the whole $\tau$-region.

In addition to the transient Fano resonance revealed in the region of $\tau \lesssim T_{\boldsymbol{q}}$ \cite{hase1}, the two more quantum-mechanical effects --- the plasmon-phonon resonance and the Rabi flopping --- are uncovered in ETR.
As long as $\tau \ge \tau_L/2$~\cite{comm1}, the induced carriers vary with respect to $\tau$ slowly enough in ETR that the plasmon mode can be created instantaneously.
Therefore, the resonance effect of concern differs from the delayed
formation of LO-phonon plasmon coupled modes observed out of ETR in undoped
GaAs~\cite{kuznetsov1, Huber}.
Further, the irregular transitory-signals manifested in ETR seen in Figs.~\ref{fig4}(a)-(c) are distinct from CA arising from interference during the overlap of the pump and probe pulses~\cite{ca}.
In particular, in the former, $Q_{\boldsymbol{q}}(\tau)$ is amplified by the formation of anticrossing due to the plasmon-LO-phonon interaction, while, in the latter, the observed signal is due to the intrinsic characters of the laser cavity.
As regards the Rabi flopping, the significance of it in CP dynamics has been overlooked thus far, though pointed out just regarding CA~\cite{lebedev}.

To conclude, the two quantum-mechanical effects of the resonant plasmon-LO-phonon interaction and the Rabi flopping are disclosed.
It is found that the underlying physics of the CP dynamics in ETR is enriched by these effects.
In particular, the former effect stands out, causing the striking cusp structures in $\Theta_{\boldsymbol{q}}(\tau)$ and $A_{\boldsymbol{q}}(\tau)$.
Due to the irregular alteration of both $\Theta_{\boldsymbol{q}}(\tau)$ and $A_{\boldsymbol{q}}(\tau)$ with respect to $\tau$, $Q_{\boldsymbol{q}}(\tau)$ shows the anomalous oscillatory pattern just observed in ETR.
It is expected that such effects are confirmed by experiments by minimizing the masking effect due to CA, for instance, using orthogonal polarizations of the pumping and probing radiation.
More practically, a combination of few-cycle laser pulses in near-infrared (or visible) region with attosecond extreme ultraviolet pulses may enable us to monitor ETR dynamics without CA~\cite{Schultze}. The quantum-mechanical effects described in the present study will be applicable for other attractive systems, such as SiC~\cite{Kato} and diamond~\cite{Ishioka}. 
On the other hand, the Rabi flopping is discernible in experiments by measuring the asymptotes of $\theta_{\boldsymbol{q}}$ and $A^0_{\boldsymbol{q}}$ as a function of $\Omega_{0cv}$ up to more than $\Omega^{(2\pi)}_{0cv}$.


\begin{acknowledgments}
This work was supported by JSPS KAKENHI Grants No. JP23540360 and No. JP15K05121.
\end{acknowledgments}


\begin{thebibliography}{00}
\bibitem{comm1}Because of a steep change of pulse envelope function $f(t)$ with respect to $t$, the adiabatic approximation supposed in Ref.~\cite{watanabe} is not always correct in $|t| \le \tau_L/2$.
\end{thebibliography}

\end{document}